\documentclass[%
reprint,
superscriptaddress,
amsmath,
amssymb,
aps,
prl,
longbibliography
]{revtex4-1}

\usepackage[T1]{fontenc}
\usepackage[utf8]{inputenc}

\usepackage{times}
\usepackage[unicode=true,breaklinks=true,colorlinks=true]{hyperref}
\hypersetup{citecolor=blue,urlcolor=blue}
\usepackage{graphicx}
\usepackage{bm}
\usepackage{bbold}
\usepackage{mathtools}
\usepackage{tensor}

\newcommand{\be}{\begin{equation}}
\newcommand{\ee}{\end{equation}}
\newcommand{\bea}{\begin{eqnarray}}
\newcommand{\eea}{\end{eqnarray}}

\newcommand{\ket}[1]{|#1\rangle}
\newcommand{\bra}[1]{\langle#1|}

\newcommand{\new}[1]{\textcolor{black}{#1}}
\newcommand{\revise}[1]{\textcolor{black}{#1}}

\usepackage{pdfpages}
\usepackage{tikz}
\makeatletter
\AtBeginDocument{\let\LS@rot\@undefined}
\makeatother

\begin{document}

\title{\new{Experimental estimation of the quantum Fisher information from randomized measurements}}
\author{Min~Yu}
\thanks{These authors contributed equally.}
\author{Dongxiao~Li}
\thanks{These authors contributed equally.}
\author{Jingcheng~Wang}
\author{Yaoming~Chu}
\author{Pengcheng~Yang}
\author{Musang~Gong}
\affiliation{School of Physics, International Joint Laboratory on Quantum Sensing and Quantum Metrology, Institute for Quantum Science and Engineering, Huazhong University of Science and Technology, Wuhan 430074, China}
\author{Nathan~Goldman}
\email{ngoldman@ulb.ac.be}
\affiliation{Center for Nonlinear Phenomena and Complex Systems, Universit\'e Libre de Bruxelles, CP 231, Campus Plaine, B-1050 Brussels, Belgium}
\author{Jianming~Cai}
\email{jianmingcai@hust.edu.cn}
\affiliation{School of Physics, International Joint Laboratory on Quantum Sensing and Quantum Metrology, Institute for Quantum Science and Engineering, Huazhong University of Science and Technology, Wuhan 430074, China}
\affiliation{Wuhan National Laboratory for Optoelectronics, Huazhong University of Science and Technology, Wuhan 430074, China}
\date{\today}
\begin{abstract}
\new{The quantum Fisher information (QFI) represents a fundamental concept in quantum physics. On the one hand, it quantifies the metrological potential of quantum states in quantum-parameter-estimation measurements. On the other hand, it is intrinsically related to the quantum geometry and multipartite entanglement of many-body systems. Here, we explore how the QFI can be estimated via randomized measurements, an approach which has the advantage of being applicable to both pure and mixed quantum states. In the latter case, our method gives access to the sub-quantum Fisher information, which sets a lower bound on the QFI. We experimentally validate this approach using two platforms:~a nitrogen-vacancy center spin in diamond and a 4-qubit state provided by a superconducting quantum computer. We further perform a numerical study on a many-body spin system to illustrate the advantage of our randomized-measurement approach in estimating multipartite entanglement, as compared to quantum state tomography.  Our results highlight the general applicability of our method to general quantum platforms, including solid-state spin systems, superconducting quantum computers and trapped ions, hence providing a versatile tool to explore the essential role of the QFI in quantum physics.}
\end{abstract}
\maketitle
\new{Quantum technologies promise appealing advantages in various practical applications. As a prime example, quantum metrology \cite{Giovannetti-2011-Nat.Pho} exploits quantum resources such as entanglement \cite{Giovannetti2006}, coherence \cite{Pires-2018-PRA}, squeezing \cite{Maccone-2020-Quantum} and criticality \cite{Zanardi-2008,Rams-2018,Chu-2021,Frerot-2018} to achieve unprecedented measurement performance. \new{This} has applications in a variety of fields, including precision measurements in physics \cite{Abbott-2016-PRL,Chin-2001-PRA,Berglund-1995-PRL}, material science \cite{Casola2018} and biology \cite{Kucsko2013}. The metrological potential of quantum resource states is quantified by the quantum Fisher information (QFI) \cite{Samuel-1996-AP}, which is an extension of the Fisher information \cite{Fisher-1925-PCPS} in the quantum realm. According to the quantum Cram\'{e}r-Rao bound, the inverse of the QFI sets a fundamental limit on the accuracy of parameter-estimation measurements~\cite{paris2009}. }
\new{Besides its role in quantum metrology}, the QFI also represents a fundamental concept in quantum physics. It has been shown to exhibit deep connections with multipartite entanglement \cite{Pezze2009,Hyllus-2012-PRA,Toth-2012-PRA,Hauke-2016-NP}. From a geometric perspective, the QFI characterizes the distinguishability between neighboring quantum states in parametric space, which is closely related to the concept of quantum geometric tensor~\cite{kolodrubetz2017geometry}. It thus plays a significant role in numerous quantum phenomena \cite{Pezze-2018-RMP}, including quantum phase transitions \cite{Zanardi-2008}, quantum Zeno dynamics \cite{Smerzi-2012-PRL}, as well as in a variety of quantum information processing protocols.
\new{Identifying experimental methods to extract the QFI of arbitrary quantum states is an outstanding challenge, which is currently under intense investigation~\cite{Strobel-2014-Science,Frowis-2016-PRL,Lu-2020-PRL,liu2020saturating}.} Indeed, the QFI is not a linear function of quantum states and it is not related to any observable that can be directly accessed~\cite{Samuel-1996-AP}. In principle, one can obtain the value of the QFI for general quantum states (pure or mixed) based on quantum state tomography. However, the number of required measurements increases rapidly with the system size, hence resulting in exceptionally heavy experimental overheads for many-body quantum systems. \new{One route is provided by dynamical susceptibilities \cite{Hauke-2016-NP}, which can be probed spectroscopically~\cite{Ozawa_Goldman_2018PRB,Ozawa_Goldman_2019PRR} and give access to the QFI of thermal states.} Although, it is possible to extract the Fisher information based on the Hellinger distance~\cite{Strobel-2014-Science}, the experimental determination of the QFI for general states still remains a challenging task \new{as it requires, by definition, the determination of the optimal measurement for which the Fisher information} is maximized \cite{Samuel-1996-AP}.
In this Letter, we \new{explore the possibility of evaluating the QFI of general quantum states, without any a priori information on the system, by adopting a method based on randomized measurements.} Our proof-of-concept experimental demonstration is performed using two independent platforms: a nitrogen-vacancy (NV) center spin in diamond and a superconducting quantum computer (IBM Q). Our scheme only relies on random measurements \new{performed} on single copies of the quantum system \cite{Zhang-2017-PRA}, and it does not require full quantum state tomography. Such a unique feature can significantly relax the experimental requirements on non-local operations or the number of measurements. It is worth noting that  techniques based on randomized measurements have been exploited in different physical contexts, such as the estimation of the $n$-th moment of general quantum states \cite{van-2012-PRL}, the R\`{e}nyi entanglement entropy \cite{Elben-2018-PRL,brydges2019probing}, the overlap of two mixed states \cite{Elben-2020-PRL-overlap}, the mixed-state entanglement \cite{Elben-2020-PRL,Zhou-2020-PRL}, and the many-body Chern number \cite{cian2020manybody}. \new{In this sense, the present scheme for the estimation of the QFI should find applications in various quantum platforms. }
We start by considering the standard scenario of quantum parameter estimation, in which the information of the parameter $\theta$ is encoded in a parameter-dependent \new{quantum} state $\rho_{\theta}$. The maximum available information to extract $\theta$ is determined by the QFI, denoted as $\mathcal{F}_{\theta}$, which quantifies the distinguishability between a general state $\rho_{\theta}$ and its neighboring state $\rho_{\theta+d\theta}$. \new{As dictated} by the quantum Cramér-Rao bound, the ultimate precision of quantum parameter estimation is given by $\delta \theta=1/\sqrt{\mathcal{F}_{\theta}}$. \new{For a general quantum state, spectrally decomposed as $\rho_\theta=\sum_\lambda p_\lambda|\lambda\rangle\langle\lambda|$, with $p_\lambda$ and $|\lambda\rangle$ its eigenvalues and corresponding eigenstates,} the QFI can be written as \cite{LiuJPA2019}
\begin{eqnarray}\label{defi_qfi}
\mathcal{F}_{\theta}(\rho_\theta)=\sum_{\lambda,\lambda'}\frac{2|\langle\lambda|\partial_\theta\rho_\theta|\lambda'\rangle|^2}{p_\lambda+p_{\lambda'}},
\end{eqnarray}
where the sum only includes the terms with $p_\lambda+p_{\lambda'}\neq0$, and $\partial_\theta\rho_\theta=\partial\rho_\theta/\partial\theta$. \new{Equation \eqref{defi_qfi} suggests} that the QFI is not directly associated with any physical observable. \new{A direct way to measure the QFI would be to fully reconstruct} the density matrices $\rho_{\theta}$ and $\rho_{\theta+d\theta}$, by performing quantum state tomography; this was demonstrated in Ref. \cite{Lu-2020-PRL}, \new{where the Bloch vector of the quantum state was measured in view of reconstructing} the QFI \cite{Zhong2013}. However, quantum state tomography requires rapidly increasing number of measurements as the system dimension grows.
\begin{figure}[t]
\centering
\includegraphics[width=9cm]{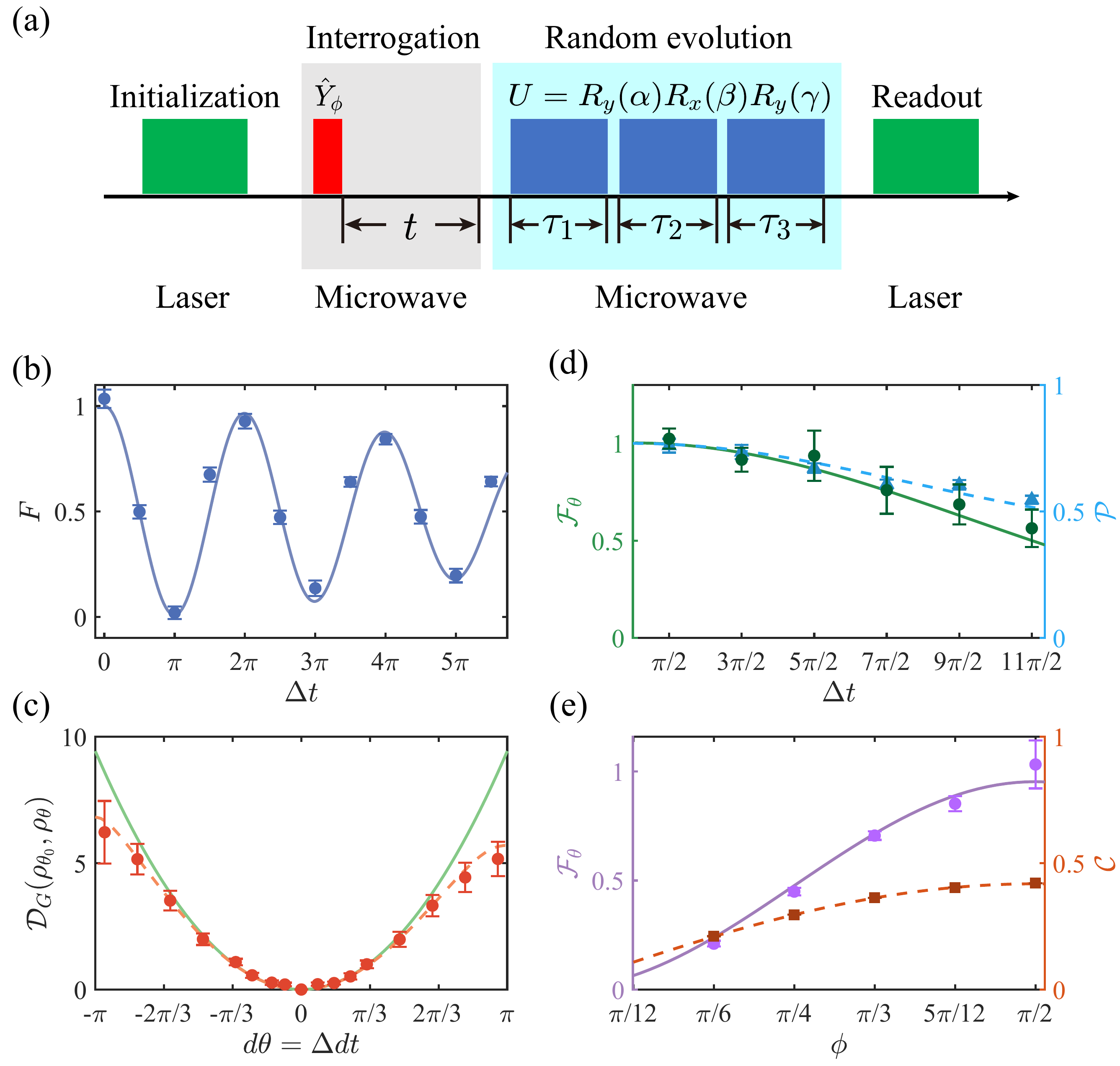}
\caption{({\bf a}) The pulse sequence for measuring the QFI in a Ramsey experiment, using an NV center spin quantum sensor in diamond. The NV center spin is initially polarized in the state $\vert 0\rangle$ by applying a green (532 nm) laser pulse and the $\phi$-dependent resource state $\vert \psi_{\phi}(0) \rangle$ is prepared via a subsequent microwave pulse $\hat Y_{\phi}$ (i.e. a rotation around the $\hat{y}$ axis by an angle $\phi$). The interrogation (i.e. the free evolution) of duration $t$ results in the final parameter-dependent quantum state $\rho_{\theta}=\rho(t)$ with $\theta=\Delta t$, see Eq.\eqref{eq:state}. The random measurement is implemented by three microwave pulses (blue), with random parameters, followed by spin-dependent fluorescence measurement. ({\bf b}) The fidelity between the evolved state $\rho(t)$ and the initial state $\vert \psi_{\phi}(0)\rangle$ obtained from randomized measurements (circle), which is compared with the result from standard deterministic projective measurement (solid curve). ({\bf c}) The modified Bures distance $\mathcal{D}_G$ between the state $\rho_{\theta_0}$ and $\rho_{\theta}$ as a function of $d\theta=\theta-\theta_0=\Delta dt$ with $\theta_0=3\pi/2$. The dashed curve is the polynomial fit to the experiment data (circles), while the solid curve represents the quadratic fit, the corresponding coefficient of which provides an estimation for the value of the QFI [see Eq.\eqref{l_qfi}]. The other experiment parameters in ({\bf b}) and ({\bf c}) include the detuning $\Delta = (2\pi)1.459$ MHz, the angle ${\phi}=\pi/2$, and the number of random measurements $n=400$. The coherence time of the NV center spin is estimated to be $T^{*}_{2} = 2.58\pm 0.2$ $\mu s$. ({\bf d}) The QFI $\mathcal{F}_{\theta}$ \revise{and the purity $\mathcal{P}$} of the evolved state in Ramsey experiment as a function of the free evolution time $t$. The noise causes decoherence and the system evolves into mixed states, which is evidenced by the decrease of the QFI. ({\bf e}) The QFI $\mathcal{F}_{\theta}$ \revise{and the coherence $\mathcal{C}=|\rho_{12}(t)|$} of the states after a free evolution time $t=3\pi/(2\Delta)$ from different initial resource states $\vert \psi_{\phi}(0)\rangle =\cos(\phi/2)\vert 0\rangle +\sin(\phi/2)\vert -1\rangle $. The detuning is $\Delta=(2\pi) 1.459$ MHz.}
\label{model}
\end{figure}

\new{Starting from the superfidelity \cite{Miszczak-2009-QIC}, defined as
\begin{equation}\label{supf}
g(\rho_1,\rho_2)=\big[{\rm Tr}(\rho_1\rho_2)+\sqrt{(1-{\rm Tr}\rho_1^2)(1-{\rm Tr}\rho_2^2)}\big],
\end{equation}
one can introduce a quantity $\mathcal{F}_{G}(\rho_\theta)$ that provides a lower bound to the QFI for any general quantum state:}
\begin{eqnarray}\label{l_qfi}
\mathcal{F}_{\theta}(\rho_\theta) \geq \mathcal{F}_{G}(\rho_\theta)\equiv  \frac{ \mathcal{D}_G(\rho_\theta,\rho_{\theta+d\theta})}{(d\theta)^2},
\end{eqnarray}
where $\mathcal{D}_G(\rho_1,\rho_2)=8 { [1-g(\rho_1,\rho_2)]}$ denotes the modified Bures distance between two quantum states $\rho_1$ and $\rho_2$ \cite{Miszczak-2009-QIC}. \new{We note that $\mathcal{F}_{G}(\rho_\theta)$ was introduced recently in \cite{Cerezo2021}, where it was coined ``sub-quantum Fisher information" (sub-QFI)}. Hence, by measuring the superfidelity, one is able to extract \new{the lower bound of the QFI for any general quantum state. Moreover, the inequality~\eqref{l_qfi} becomes an equality for pure states, so that the exact QFI $\mathcal{F}_{\theta}$ of a pure quantum state can be obtained through the superfidelity}.
\new{In our experimental study, we first measure the QFI of quantum sensor states encoded by a single nitrogen-vacancy (NV) center spin in diamond \cite{Balasubramanian2008,Maze2008,Rondin2014}. We measure the QFI within the scenario of quantum parameter estimation, provided by standard Ramsey interferometry.} By applying an external magnetic field $B_z$ along the NV axis, we lift the degeneracy of the spin states $m_s=\pm 1$ of the NV center and encode a qubit using the two spin sublevels $m_s=0,-1$ with an energy gap $\omega_0=D-\gamma_e B_z$, where $D=(2\pi)2.87$ GHz is the zero-field splitting and $\gamma_e$ is the electronic gyromagnetic ratio. The external magnetic field is chosen as $B_z\simeq 510$ G such that the associated nitrogen nuclear spin in proximity to the NV center is polarized by optical pumping. The measurement protocol in our experiment is shown in Fig.\ref{model}(a). The NV center spin is prepared in the state $\vert \psi_{\phi}(0)\rangle =\cos(\phi/2)\vert 0\rangle +\sin(\phi/2)\vert -1\rangle $ by applying a microwave pulse $\hat{Y}_{\phi}$ with frequency $\omega$ to rotate the spin around the $\hat{y}$ axis by an angle $\phi$ after the optical initialization. The NV center spin evolves as  \begin{equation}
\rho(t)=\left[ \begin{array}{cc}
    \cos^2(\phi/2) & \frac{1}{2}\sin(\phi)e^{i\theta-(t/T_2^\ast)^2} \\
    \frac{1}{2}\sin(\phi)e^{-i \theta -(t/T_2^\ast)^2}
     & \sin^2(\phi/2)
\end{array} \right],
\label{eq:state}
\end{equation}
where $\theta\!=\!\Delta t$ with $\Delta\!=\!\omega-(D-\gamma_eB_z)$ is the unknown parameter, from which one can infer the accurate value of the magnetic field $B_z$. The dephasing noise causes the decay of the NV center spin coherence at a rate of $(T_2^\ast)^{-1}$. In our experiment, the randomized measurement is realized by a spin-dependent fluorescence measurement following three microwave pulses, which implement a random unitary transformation $U$ of the form [Fig.\ref{model}(a)]
\begin{eqnarray}\label{rU}
U=R_x(\alpha)R_y(\beta)R_x(\gamma),
\end{eqnarray}
where $R_{v}(\varphi)$ denotes the rotation around the $\hat{v}$ axis by an angle $\varphi$. The rotation angles $\alpha$, $\beta$, $\gamma$ are chosen in a random way such that the unitary transformation $U$ acting on the NV center spin \new{belongs to the circular unitary ensemble} \cite{Elben-2018-PRL}. The subsequent fluorescence measurement is equivalent to the projective measurement on the state $\rho(t)$  along a random basis $\ket{u}=U^{\dagger}\vert 0 \rangle$, the outcome of which is given by $p(t)=\mbox{Tr}(\vert 0\rangle \langle 0\vert U\rho(t)U^{\dagger})=\bra{u}\rho(t)\ket{u}$.  We remark that the principle demonstrated in this work holds not only for \new{local but also for nonlocal random unitary transformations}. It is also worth pointing out that the measurement can be generalized from a single projective basis to a collective observable suitable for many-body systems \cite{supp}, \new{such as cold atoms and spin ensembles}.

From the random measurements, we can obtain the fidelity of the quantum state $\rho(t)$ with respect to the initial state as $F(t)=\mbox{Tr}(\rho(t)\rho(0))=6 \langle  p(t)p(0) \rangle -1$ \cite{supp}, where $\langle$-$\rangle$ denotes the average over $n$ random matrices $U$. The result is shown in Fig.~\ref{model}(b), which shows a good agreement with the standard deterministic projective measurement. We proceed by performing local measurements on the states $\rho_{\theta}\!=\!\rho(t)$ and $\rho_{\theta+d\theta}\!=\!\rho(t+dt)$ with $d\theta\!=\!\Delta d t$, which allows us to obtain the values of $\mbox{Tr}(\rho_{\theta}\rho_{\theta+d\theta})$,  $\mbox{Tr}(\rho_{\theta}^2)$ and $\mbox{Tr}(\rho_{\theta+d\theta}^2)$ \cite{van-2012-PRL,Elben-2018-PRL,brydges2019probing,Elben-2020-PRL-overlap}; using Eq.\eqref{supf}, this eventually provides the modified Bures distance $\mathcal{D}_G(\rho_\theta,\rho_{\theta+d\theta})$.  The polynomial fit to $\mathcal{D}_G(\rho_\theta,\rho_{\theta+d\theta})$ is shown as the dashed line in Fig.\ref{model}(c), while the solid line only includes  the quadratic term. The good agreement between the two curves indicates the suitable range of $d\theta$ within which the QFI can be extracted from the coefficient of the quadratic term. According to the Taylor expansion of the superfidelity between the state $\rho_{\theta}$ and $\rho_{\theta+d\theta}$, namely  $\mathcal{D}_G(\rho_\theta,\rho_{\theta+d\theta})=\mathcal{F}_G(\rho_\theta)d\theta^2+\mathcal{O}(d\theta^3)$ \cite{supp}, we can extract \new{$\mathcal{F}_G(\rho_\theta)$ from the coefficient of the quadratic term. \revise{In the present single-qubit case, the superfidelity is equivalent to the Uhlmann-Jozsa fidelity for both pure and mixed states~\cite{Miszczak-2009-QIC}; we thus obtain the exact QFI through  $\mathcal{F}_{\theta}=\mathcal{F}_G(\rho_\theta)$; see Eq.\eqref{l_qfi}}.}
\begin{figure}[t]
\centering
\includegraphics[width=82mm]{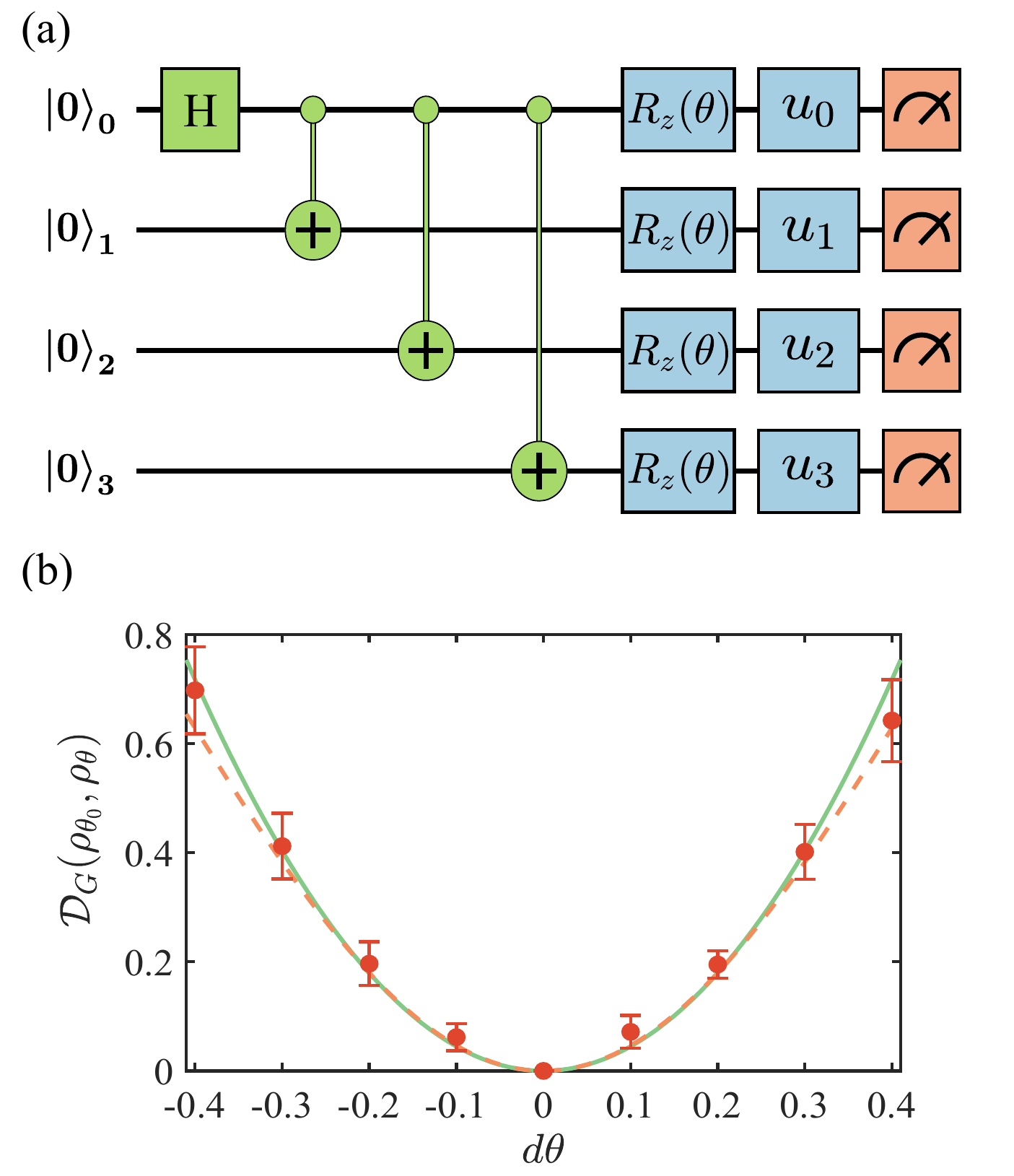}
\caption{({\bf a}) Quantum circuit for the measurement of the QFI for a four-qubit entangled state. The Hadamard gate and the control-NOT gates are used to prepare GHZ state, which is used for the estimation of the parameter $\theta$ as induced by the rotation $R_z(\theta)$. Local random matrices $u_j$ are applied to realize randomized measurements. ({\bf b}) The modified Bures distance $\mathcal{D}_G$ between the state $\rho_{\theta_0}$ and $\rho_{\theta}$ as a function of $d\theta\!=\!\theta-\theta_0$; here we set $\theta_0\!=\!0$. The dashed curve is the polynomial fit to the experiment data (circles) generated by the IBM Quantum Falcon Processor (ibmq$\_$belem v1.0.3), while the solid curve represents the quadratic fit. \revise{The experimental sub-QFI $\mathcal{F}_{G}=4.301\pm0.22$ is compatible with the \new{value $\mathcal{F}_{G}=4.4807$ obtained from quantum state tomography \cite{supp}. This result provides a lower bound for the exact QFI}, which is $\mathcal{F}_{\theta}=6.5147$ in this case~\cite{supp}}. The number of randomized measurements is $n\!=\!400$.}
\label{GHZ}
\end{figure}
We show in Fig.\ref{model}(d) the measured dynamical evolution of the QFI for the state $\rho(t)$. The result shows \new{the influence of dephasing noise} on the QFI and thereby the metrological potential of the evolving state. We note that the evolution of the QFI may serve as an indicator of the environmental noise properties in open quantum system, e.g.~non-Markovianity \cite{Lu-2020-PRL}. The present technique does not require \new{any prior information on the} system and thus provides a powerful tool to investigate general open quantum systems using the concept of the QFI. Furthermore, we measure the QFI \revise{of the evolved state $\rho(t)$ for different initial states} $\vert \psi_{\phi}(0)\rangle =\cos(\phi/2)\vert 0\rangle +\sin(\phi/2)\vert -1\rangle $, to establish the relation between the QFI and the coherence of the resource states \new{(as quantified by the off-diagonal element of the density matrix  $\mathcal{C}=|\rho_{12}(t)|$). From the results displayed in Fig.\ref{model}(e), we find that $\rho(t)$ is indeed mixed due to the influence of noise, and that both the QFI and the coherence increase as the initial angle $\phi$ is increased towards $\pi/2$}. The experimental data agrees well with the exact value of the QFI, which demonstrates the validity of the present scheme in measuring the QFI for both pure and mixed quantum states.
\begin{figure}[t]
\centering
\includegraphics[width=78mm]{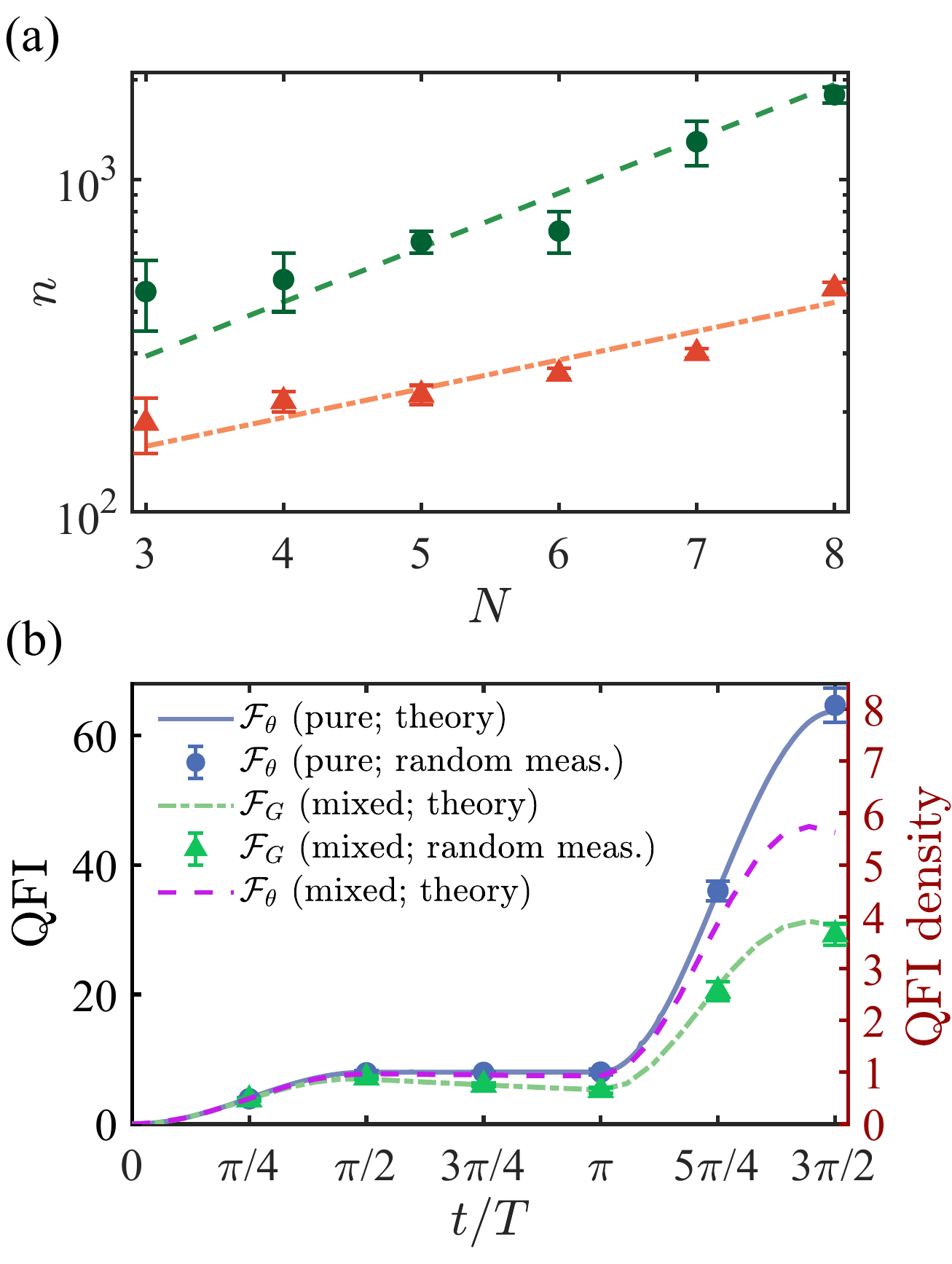}
\caption{({\bf a}) \new{The number of measurements $n$ required to achieve an average relative error smaller than $9\%$ on the QFI (circles, $\gamma\!=\!0$) or on the sub-QFI (triangle, $\gamma/g\!=\!0.01$),  as a function of the number of ions. The dashed and dash-dotted lines are the corresponding exponential fits $2^{a+b N}$ with $a=6.57\pm1.40,b=0.54\pm0.19$ and $a=6.44\pm1.00,b=0.29\pm0.15$, respectively. The chosen parameters are $\alpha\!=\!1.5$, $K\!=\!20$, $d\theta\!=\!0.1$, and $g\!=\!\Omega\!=\!\delta\!=\!1/T$. ({\bf b}) Numerical simulation of the QFI ($\gamma=0$; full line) and sub-QFI ($\gamma/g=0.01$; dash-dotted line), as a function of the evolution time, for a 8-qubit GHZ state; the QFI of the mixed state is displayed as a dashed line. The circles and triangles are the results obtained from randomized measurements.  The number of randomized measurements is $n\!=\!1000$, and other parameters are the same as in ({\bf a}).} }
\label{scaling}
\end{figure}

%
As a second demonstration, we generate a 4-qubit GHZ state (\new{with estimated fidelity of} $78\%$ \cite{supp}) and we estimate the QFI from randomized measurements using the Qiskit by IBM Quantum Experience; the corresponding quantum circuit is shown in Fig.~\ref{GHZ}(a).  To estimate the lower bound of the QFI for the parameterized state $\rho_\theta=e^{-i\theta J_z}\rho e^{i\theta J_z}$, where  $J_z=\sum\limits_{j=1}^4\sigma_z^{(j)}/2$, we apply local random unitary transformations $u_j$ from the classical compact group, where $u_j$ is determined by the random parameters $\lambda_j,\theta_j,\varphi_j$ through
\begin{eqnarray}\label{l_qfi_2}
u_j=\left[\begin{array}{cc}
\cos(\frac{\theta_j}{2}) & -e^{i\lambda_j}\sin(\frac{\theta_j}{2})
\\
e^{i\varphi_j}\sin(\frac{\theta_j}{2}) & e^{i(\varphi_j+\lambda_j)}\cos(\frac{\theta_j}{2})
\end{array}
 \right].
\end{eqnarray}
We obtain the modified Bures distance $\mathcal{D}_G(\rho_{\theta_0},\rho_{\theta})$, and the polynomial fit is shown as the dashed line in Fig.~\ref{GHZ}(b). The coefficient of the quadratic term \new{provides the sub-QFI, $\mathcal{F}_{G}(\rho_{\theta_0})=4.301\pm0.22$,} which is in good agreement \new{with the value $\mathcal{F}_{G}=4.4807$ obtained from quantum state tomography \cite{supp}. The measured $\mathcal{F}_{G}$ provides the lower bound of the exact QFI, which is $\mathcal{F}_{\theta}=6.5147$ in this case~\cite{supp}.}
\new{We now explore the applicability of our QFI measurement in the context of many-body quantum physics \cite{Pezze2009}, setting the focus on entanglement dynamics in an open $N$-qubit system. We again consider a GHZ state \cite{leibfried2005creation}, obtained by applying the unitary operation  $U_N\!=\!\exp(i{\pi}J_x/2)\exp(i{\pi}J_z^2/2)\exp(i{\pi}J_x/2)$ on an initialized state $|00\dots0\rangle$; here $J_\alpha=\sum_{j=1}^N\sigma^{(j)}_{\alpha}/2$, where $\sigma_\alpha^{(j)}$ is the Pauli matrix along the $\alpha$ direction of the $j$-th qubit. In order to simulate the noise that always affects real experiments, we describe the dynamical evolution of the system by a quantum master equation, whose dissipator is taken in the form}
\begin{equation}
\mathcal{L}\rho=\sum_{j=1}^N L_j\rho L_j^\dag-\frac{1}{2}(L_j^\dag L_j\rho+\rho L_j^\dag L_j),
\end{equation}
where the Lindblad operator $L_j=\sqrt{\gamma}\sigma^{(j)}_z$ describes the dephasing of the $j$-th qubit with a dephasing rate $\gamma$. To measure the \new{sub-QFI of the mixed state} $\rho_\theta=e^{-i\theta J_z}\rho e^{i\theta J_z}$, we design random matrices through the time-evolution operator $U=\prod\limits_{m=1}^{K} e^{-i \mathcal{H}_{m} T}$, where $\mathcal{H}_{m}=\sum_{j=1}^{N} \Delta_{m}^{(j)}\sigma^{(j)}_z+H_s$. The first term $\Delta_{m}^{(j)}\sigma^{(j)}_z$ represents an on-site disorder drawn from a normal distribution with the standard deviation $\delta$, while the second term $H_s$ denotes the Ising Hamiltonian that can be implemented in trapped-ion systems~\cite{Zhang-2017-Nature,Jurcevic-2017-PRL}, namely $H_s=\sum\limits_{(k<l)}g|k-l|^{-\alpha}\sigma^{(k)}_x\sigma^{(l)}_x+\Omega\sum\limits_{k=1}^N \sigma^{(k)}_x$, with $0<\alpha<3$ and $g$ the nearest-neighbor coupling.
\new{Figure \ref{scaling}(a) shows the scaling of the number of randomized measurements that are required to achieve an average relative error on the QFI that is smaller than some threshold $\epsilon\!=\!9\%$. In the absence of noise ($\gamma\!=\!0$), the system is in a pure state, and the number of required measurements is shown to scale as $\sim 2^{(0.54\pm0.19) N}$ (dashed line). We remark that such a scaling represents a significant improvement over quantum state tomography, for} which the number of required measurements scales as $\sim 2^{2 N}$ \cite{Gross-2010-PRL}. The improvement becomes even more pronounced upon considering the mixed states generated by noise. In this case, the number of required measurements scales as $~2^{(0.29\pm0.15) N}$ (dash-dotted line). These results illustrate how the fluctuations across the unitary ensemble decrease as the system becomes highly mixed \cite{brydges2019probing}.

\new{Importantly, the QFI is known to serve as a witness of multipartite entanglement~\cite{Hyllus-2012-PRA,Toth-2012-PRA}:~a general quantum state $\rho$ must be $(m+1)$-partite entangled if the QFI density $\mathcal{F}_{\theta}/N>m$.} \new{In Fig.~\ref{scaling}(b), we show the QFI (resp.~sub-QFI) for pure (resp.~mixed) states, together with the corresponding density $\mathcal{F}_{\theta}/N$ as a function of the evolution time.  These results demonstrate the efficiency of our scheme in measuring the lower bound of the QFI, and hence the dynamics of multipartite entanglement, in open many-body  quantum systems.}
{\it Conclusions \& outlook.---} \new{To summarize, we used several quantum platforms to explore an efficient scheme to estimate the QFI of general quantum states based on randomized measurements.  Our method does not rely on full quantum state tomography, and in fact, it exhibits significantly enhanced efficiency over quantum state tomography when applied to many-body quantum systems. The extension of our work to other quantum systems may provide powerful tools to estimate the QFI in different contexts, allowing for the experimental exploration of the QFI in various quantum phenomena.}
{\it Acknowledgements.---} \new{We thank Markus Heyl and Tomoki Ozawa for valuable discussions}. This work is supported by National Natural Science Foundation of China (Grant No.~11874024, 11690032, 12047525), the Open Project Program of Wuhan National Laboratory for Optoelectronics (No. 2019WNLOKF002), and the Fundamental Research Funds for the Central Universities. N.G. is supported by the FRS-FNRS (Belgium) and the ERC Starting Grant TopoCold.

\bibliography{reference_v2}
\onecolumngrid
 

\section*{Supplementary material}

## Supplementary Material

### S.1. Extract quantum state information via general random measurements

The key for the estimation of the QFI in the main text is the measurement of quantum state purity and overlap in order to obtain the superfidelity. Given a random unitary matrix $U$ in the Hilbert space $\mathcal{H}_N$ with the dimension $N$, we can obtain the identity

$$UU^\dagger = 1, \tag{S.1}$$

which means that the $m$-th diagonal term of $UU^\dagger$ is

$$\sum_{n=1}^{N} U_{mn} U^*_{mn} = 1. \tag{S.2}$$

If $U$ is a random matrix distributed according to the Haar measure, i.e. chosen in the Circular Unitary Ensemble (CUE), one can prove that [1]

$$\langle U_{mn} U^*_{mn} \rangle = 1/N, \tag{S.3}$$

which is a special case of the general result

$$\langle U_{kl} U^*_{mn} \rangle = \delta_{km} \delta_{ln}/N. \tag{S.4}$$

Here $U_{kl}$ and $U^*_{mn}$ are the matrix elements of $U$ and the corresponding complex conjugate respectively. $\langle \cdot \rangle$ denotes the average over the distribution of random unitaries. Higher-order averages follow from the Isserlis ("Gaussian-moments") theorem [1–3] including

$$\langle U_{mk} U^*_{nk} U_{m'k'} U^*_{n'k'} \rangle = \frac{\delta_{mn}\delta_{m'n'} + \delta_{kk'}\delta_{mn'}\delta_{m'n}}{N^2 - 1} - \frac{\delta_{mn'}\delta_{m'n} + \delta_{kk'}\delta_{mn}\delta_{m'n'}}{N(N^2 - 1)}. \tag{S.5}$$

We stress that the present derivation is valid for random unitary matrices $U$, including local [4, 5] and nonlocal unitaries.

Random measurements can be implemented by first applying the random matrix to a density matrix $\rho_1$ with the same dimension $N$, which transforms the state as follows

$$\mathcal{U}(\rho_1) = U\rho_1 U^\dagger = \sum_{m,n,l,l'} \rho^{(1)}_{mn} U_{lm} U^*_{l'n} |l\rangle\langle l'|, \tag{S.6}$$

where $\rho^{(1)}_{mn}$ is the element of $\rho_1$, i.e. $\rho^{(1)}_{mn} = \langle m|\rho_1|n\rangle$, and $\{|n\rangle\}$ forms a basis. The subsequent measurement in a basis state $|k\rangle \in \mathcal{H}_N$ gives the probability $P_1(k)$ as

$$P_1(k) = \text{Tr}(|k\rangle\langle k|\mathcal{U}(\rho_1)) = \sum_{m,n} \rho^{(1)}_{mn} U_{km} U^*_{kn}, \tag{S.7}$$

the corresponding average of which is thus given by (according to Eq. (S.4))

$$\langle P_1(k) \rangle = \sum_{m,n} \rho^{(1)}_{mn} \langle U_{mk} U^*_{nk} \rangle = \sum_{m,n} \rho^{(1)}_{mn} \delta_{mn}/N = 1/N, \tag{S.8}$$

where we have used the identity $\text{Tr}(\rho) = \sum_m \rho_{mm} = 1$. Similarly, the probability $P_2(k)$ can be obtained by the same random matrix $U$ acting on another density matrix $\rho_2$ as

$$P_2(k) = \sum_{m,n} \rho^{(2)}_{mn} U_{km} U^*_{kn}. \tag{S.9}$$

In order to obtain the overlap between $\rho_1$ and $\rho_2$, i.e. $\text{Tr}[\rho_1\rho_2]$, we further investigate the correlation between $P_1(k)$ and $P_2(k)$

$$\langle P_1(k) P_2(k) \rangle = \sum_{m,n,m',n'} \rho^{(1)}_{mn} \rho^{(2)}_{m'n'} \langle U_{km} U^*_{kn} U_{km'} U^*_{kn'} \rangle. \tag{S.10}$$

By making use of the Isserlis theorem in Eq. (S.5), this correlation is found to be

$$\langle P_1(k) P_2(k) \rangle = \frac{\sum_{m,n} \rho^{(1)}_{mm}\rho^{(2)}_{nn} + \sum_{m,n} \rho^{(1)}_{mn}\rho^{(2)}_{nm}}{N(N+1)} = \frac{1+\text{Tr}(\rho_1\rho_2)}{N(N+1)}. \tag{S.11}$$

which leads to the following result

$$\text{Tr}(\rho_1\rho_2) = N(N+1)\langle P_1(k) P_2(k) \rangle - 1. \tag{S.12}$$

Furthermore, one can replace $N$ in the above equation (S.12) by the inverse of the average of the probability $1/P_{1(2)}(k)$ to make the result more accurate when $N$ is a finite value, see Ref. [1]. We note that the result gives the state purity if we choose $\rho_2 = \rho_1$.

On the other hand, in the above discussion we assume that the probability $P_1(k) = \text{Tr}(|k\rangle\langle k|\mathcal{U}(\rho_1))$ is measured in a single basis state $|k\rangle$, where $|k\rangle$ is an arbitrary $N$-dimensional basis of the Hilbert space $\mathcal{H}_N$ [1, 3–9]. It is worth noting that for a system consisting of $K$ $d$-dimensional objects (i.e., $N = d^K$), it may be more feasible to measure the probability $P_1(\hat{s}_\alpha)$ corresponding to the collective observable $\hat{s}_\alpha = \sum_{j=1}^{K} |\alpha\rangle_{jj}\langle\alpha|/K$, where $|\alpha\rangle_j$ denotes an arbitrary $d$-dimensional basis state for the $j$th particle. In this case, we have

$$P_1(\hat{s}_\alpha) = \frac{1}{K}\sum_{j=1}^{K} \text{Tr}(|\alpha\rangle_{jj}\langle\alpha|\mathcal{U}(\rho_1)), \tag{S.13}$$

where $\mathcal{U}(\rho_1)$ is a unitarily transformed $N$-dimensional density matrix $\rho_1$ with a random matrix $U$. For the total system, there will be $N_A^m = K!(d-1)^{K-m}/[m!(K-m)!]$ eigenstates, in which $m$ particles are in the state $|\alpha\rangle$. Therefore, the probability $P_1(\hat{s}_\alpha)$ can be expanded as

$$P_1(\hat{s}_\alpha) = \frac{1}{K}\sum_{m=1}^{K}\sum_{l_m=1}^{N_A^m} m\,\text{Tr}(|m,l_m\rangle\langle m,l_m|\mathcal{U}(\rho_1)). \tag{S.14}$$

By making use of the Eq. (S.8), the average of $P_1(\hat{s}_\alpha)$ over the random matrix is given by

$$\langle P_1(\hat{s}_\alpha) \rangle = \frac{1}{KN}\sum_{m=1}^{K} m N_A^m, \tag{S.15}$$

which means the dimension of the total system can be experimentally obtained by $N = \sum_{m=1}^{K} m N_A^m /(K\langle P_1(\hat{s}_\alpha)\rangle)$. Then we consider the same random measurement acting on another density matrix $\rho_2$ and obtain the probability $P_2(\hat{s}_\alpha)$. To calculate the overlap between $\rho_1$ and $\rho_2$, we can multiply $P_1(\hat{s}_\alpha)$ by $P_2(\hat{s}_\alpha)$

$$P_1(\hat{s}_\alpha)P_2(\hat{s}_\alpha) = \frac{1}{K^2}\sum_{m,m'}\sum_{l_m,l_{m'}} mm' \text{Tr}(|m,l_m\rangle\langle m,l_m|\mathcal{U}(\rho_1))\text{Tr}(|m',l_{m'}\rangle\langle m,l_{m'}|\mathcal{U}(\rho_2)). \tag{S.16}$$

Based on the Eq. (S.5), the corresponding average can be obtained as

$$\langle P_1(\hat{s}_\alpha)P_2(\hat{s}_\alpha)\rangle = C_1 \frac{1+\text{Tr}(\rho_1\rho_2)}{N(N+1)} + C_2 \frac{N-\text{Tr}(\rho_1\rho_2)}{N(N^2-1)},$$

where

$$C_1 = \frac{1}{K^2}\sum_m m^2 N_A^m$$

$$C_2 = \frac{1}{K^2}\sum_m \left[m^2 N_A^m(N_A^m - 1) + \sum_{m'\neq m} mm' N_A^m N_A^{m'}\right].$$

This leads to the following inverse formula to calculate the overlap between $\rho_1$ and $\rho_2$

$$\text{Tr}(\rho_1\rho_2) = \frac{N(N^2-1)\langle P_1(\hat{s}_\alpha)P_2(\hat{s}_\alpha)\rangle + C_1 - (C_1+C_2)N}{C_1 N - (C_1+C_2)}. \tag{S.17}$$

In order to verify the above result, we choose the collective observable

$$\hat{s}_0 = \sum_{j=1}^{8} |0\rangle_{jj}\langle 0|/8, \tag{S.18}$$

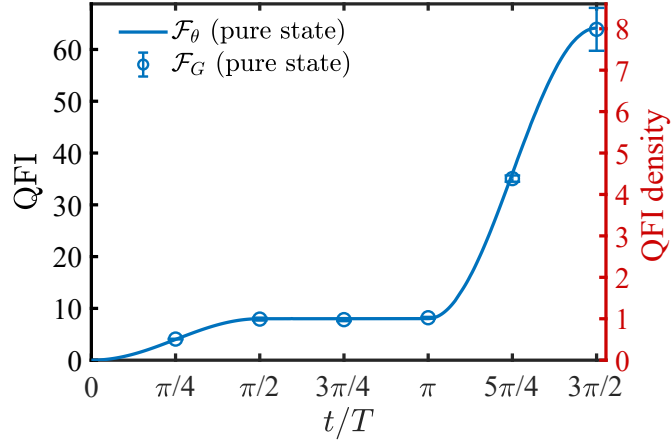

FIG. S1. The QFI and the QFI density obtained from the collective observable $\hat{s}_0 = \sum_{j=1}^{8} |0\rangle_{jj}\langle 0|/8$ (see Eq.(S.17)) as functions of the evolution time for the same 8-ion system as that of Fig.3(b) in the main text. The relevant parameters are chosen as $\gamma = 0$, $d\theta = 0.1$, $n = 2000$, $K = 20$, $\alpha = 1.5$, and $g = \Omega = \delta = 1/T$.

and use Eq.(2-3) in the main text to simulate the QFI and the QFI density for an 8-ion system in Fig.S1. The parameter-dependent state $\rho_\theta$ and the relevant parameters are the same as those of Fig.3(b) in the main text with $\gamma = 0$. The solid line is the theoretical prediction and the circles are the corresponding results obtained via random measurements based on the collective observable $\hat{s}_\alpha$. The good agreement between the theoretical description and the random measurement outcomes verifies the validity of our analytical derivation.

### S.2. Derivation of the relation between the QFI and Bures metric

For completeness, here we briefly reproduce some results of Ref. [10] that are essential to understand our scheme. The relation between the QFI and the Bures metric is

$$\mathcal{D}_B^2(\rho_\theta, \rho_{\theta+d\theta})/(d\theta)^2 s \simeq \frac{1}{4}\mathcal{F}_\theta(\rho_\theta), \tag{S.19}$$

where $\mathcal{D}_B^2(\rho_\theta, \rho_{\theta+d\theta}) = 2[1 - f(\rho_\theta, \rho_{\theta+d\theta})]$ is the Bures metric and $f(\rho_\theta, \rho_{\theta+d\theta}) = \text{Tr}(\sqrt{\sqrt{\rho_\theta}\rho_{\theta+d\theta}\sqrt{\rho_\theta}})$ is the Uhlmann-Jozsa fidelity. To prove the above identity, we consider the Taylor expansion

$$\rho_{\theta+d\theta} \simeq \rho_\theta + \frac{\partial \rho_\theta}{\partial \theta}d\theta + \frac{1}{2}\frac{\partial^2 \rho_\theta}{\partial \theta^2}(d\theta)^2, \tag{S.20}$$

which leads to

$$\sqrt{\rho_\theta}\rho_{\theta+d\theta}\sqrt{\rho_\theta} \simeq \rho_\theta^2 + \sqrt{\rho_\theta}\frac{\partial \rho_\theta}{\partial \theta}\sqrt{\rho_\theta}d\theta + \sqrt{\rho_\theta}\frac{\partial^2 \rho_\theta}{2\partial \theta^2}\sqrt{\rho_\theta}(d\theta)^2. \tag{S.21}$$

We can assume

$$\sqrt{\sqrt{\rho_\theta}\rho_{\theta+d\theta}\sqrt{\rho_\theta}} \simeq \rho_\theta + Wd\theta + Y(d\theta)^2, \tag{S.22}$$

from which the Bures metric can be rewritten as $\mathcal{D}_B^2 = -2\text{Tr}(W)d\theta - 2\text{Tr}(Y)(d\theta)^2$. By comparing the square of Eq. (S.22) with Eq. (S.21), we can find

$$\sqrt{\rho_\theta}\frac{\partial \rho_\theta}{\partial \theta}\sqrt{\rho_\theta} = \rho_\theta W + W\rho_\theta, \tag{S.23}$$

$$\sqrt{\rho_\theta}\frac{\partial^2 \rho_\theta}{2\partial \theta^2}\sqrt{\rho_\theta} = \rho_\theta Y + Y\rho_\theta + W^2. \tag{S.24}$$

By plugging $\rho_\theta = \sum_i p_{\lambda_i}|\lambda_i\rangle\langle\lambda_i|$ and $\partial_\theta \rho_\theta = \sum_i \partial_\theta p_{\lambda_i}|\lambda_i\rangle\langle\lambda_i| + \lambda_i \partial_\theta(|\lambda_i\rangle)\langle\lambda_i| + p_{\lambda_i}|\lambda_i\rangle\partial_\theta\langle\lambda_i|$ (where $\partial_\theta = \partial/\partial \theta$) into the Eq. (S.23), we can obtain the matrix element of $W$ as follows

$$[W]_{ij} = \frac{[\sqrt{\rho_\theta}\partial_\theta \rho_\theta \sqrt{\rho_\theta}]_{ij}}{p_{\lambda_i} + p_{\lambda_j}} = \frac{1}{2}\partial_\theta p_{\lambda_i}\delta_{ij} - \frac{\sqrt{p_{\lambda_i}p_{\lambda_j}}(p_{\lambda_i} - p_{\lambda_j})}{p_{\lambda_i} + p_{\lambda_j}}\langle\lambda_i|\partial_\theta|\lambda_j\rangle. \tag{S.25}$$

and thus $\text{Tr}(W) = \partial_\theta \sum_i \lambda_i = 0$. Similarly, one can derive that

$$\text{Tr}(Y) = -\sum_i \left[ \frac{(\partial_\theta p_{\lambda_i})^2}{8 p_{\lambda_i}} + 4\lambda_i \langle \partial_\theta \lambda_i | \partial_\theta \lambda_i \rangle - \sum_j \frac{8 p_{\lambda_i} p_{\lambda_j}}{p_{\lambda_i} + p_{\lambda_j}} |\langle \lambda_i | \partial_\theta \lambda_j \rangle|^2 \right] = -\frac{1}{8} \mathcal{F}_\theta, \tag{S.26}$$

where $|\partial_\theta \lambda\rangle = \partial_\theta |\lambda\rangle$, and $\mathcal{F}_\theta$ is given in Eq.(1) in the main text. Therefore, we obtain $\mathcal{D}_B(\rho_\theta, \rho_{\theta+d\theta})^2/(d\theta)^2 = -2\text{Tr}(Y) = \mathcal{F}_\theta/4$, i.e. the identity in Eq. (S.19). Since $f(\rho_\theta, \rho_{\theta+d\theta})$ is upper bounded by the superfidelity $g(\rho_\theta, \rho_{\theta+d\theta})$, we can thus obtain the lower bound of the QFI (Eq.4 in the main text)

$$\mathcal{F}_\theta(\rho_\theta) \geq \mathcal{F}_G(\rho_\theta) \equiv \frac{\mathcal{D}_G(\rho_\theta, \rho_{\theta+d\theta})}{(d\theta)^2}, \tag{S.27}$$

where $\mathcal{D}_G(\rho_\theta, \rho_{\theta+d\theta}) \equiv 8[1 - g(\rho_\theta, \rho_{\theta+d\theta})]$ is the modified Bures distance [11]. We remark that the inequality can be saturated when one of the two states is a pure state or when the dimension of the Hilbert space is 2. Consequently, $\mathcal{F}_G(\rho_\theta)$ is equivalent to the exact value of the QFI in the particular case of a single-qubit system.

### S.3. Additional experimental data

In Fig. S2, we also show the modified Bures distance $\mathcal{D}_G(\rho_{\theta_0}, \rho_\theta)$ as functions of $d\theta$ for different values of $\theta_0$. The solid, dashed, dash-dotted curves represent the corresponding quadratic fit of the experiment data, the coefficients of which provide estimations for the QFI. Moreover, the values of the QFI that are extracted in this way correspond to the circles of the Fig. 2(a) in the main text.

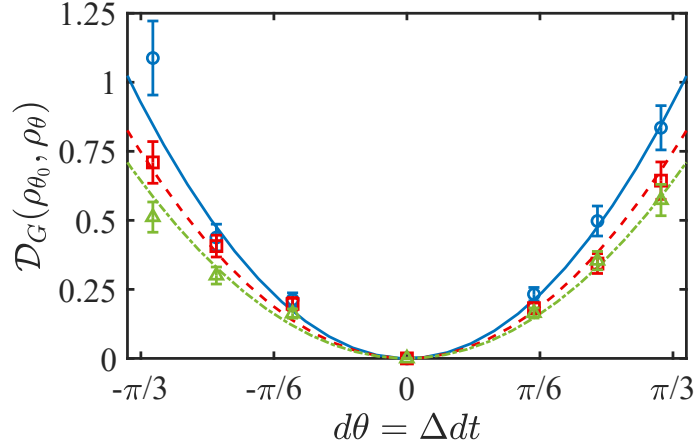

FIG. S2. The modified Bures distance $\mathcal{D}_G(\rho_{\theta_0}, \rho_\theta)$ between the state $\rho_{\theta_0}$ and $\rho_\theta$ as a function of $d\theta = \theta - \theta_0 = \Delta dt$ for $\theta_0 = 5\pi/2$ (circles), $9\pi/2$ (squares), and $11\pi/2$ (triangles). The solid, dashed, dash-dotted curves are the quadratic fit of the experiment data, the corresponding coefficients of which provide the values of the QFI as $0.94 \pm 0.13$, $0.69 \pm 0.10$, and $0.56 \pm 0.10$. The other relevant experimental parameters are the detuning $\Delta = (2\pi)1.459$ MHz, the angle $\phi = \pi/2$, and the number of random measurements is $n = 400$. The coherence time of the NV center spin is estimated to be $T_2^* = 2.58 \pm 0.2$ $\mu s$. We note that the experimental operations are the same as those of Fig.1 in the main text.

We generate a 4-qubit GHZ state with the IBM Quantum Falcon Processor (ibmq_belem v1.0.3) using the circuit as shown in Fig.2(a) of the main text. Due to the imperfection of the gates, the prepared state is not an ideal pure GHZ state. Instead, the state is a mixed state. We perform full quantum state tomography, and the entries of the corresponding density matrix that we obtain are shown in Fig.S3. This allows us to estimate that the fidelity of the prepared state with respect to the 4-qubit GHZ state reaches 78%. Based on the density matrix obtained from quantum state tomography, we calculate the parameter-dependent evolution $\rho_{\theta_0} = e^{-i\theta_0 J_z} \rho e^{i\theta_0 J_z}$ and $\rho_\theta = e^{-i\theta J_z} \rho e^{i\theta J_z}$, where $J_z = \sum_{j=1}^{4} \sigma_z^{(j)}/2$ and $d\theta = \theta - \theta_0$. Thus, we can obtain the modified Bures distance $\mathcal{D}_G(\rho_{\theta_0}, \rho_\theta)$ with $d\theta = 0, \pm 0.001, \pm 0.002, \pm 0.003$ and perform the quadratic fit of $\mathcal{D}_G(\rho_{\theta_0}, \rho_\theta) \simeq \mathcal{F}_G(\rho_{\theta_0}) d\theta^2$ to get the theoretical value of the lower bound of the QFI (following Eq.2-3 in the main text, i.e. $\mathcal{F}_G(\rho_{\theta_0})$). Similarly, by calculating the Uhlmann-Jozsa fidelity between $\rho_{\theta_0}$ and $\rho_\theta$ we can obtain the exact value of the QFI $\mathcal{F}_\theta = 6.5147$.

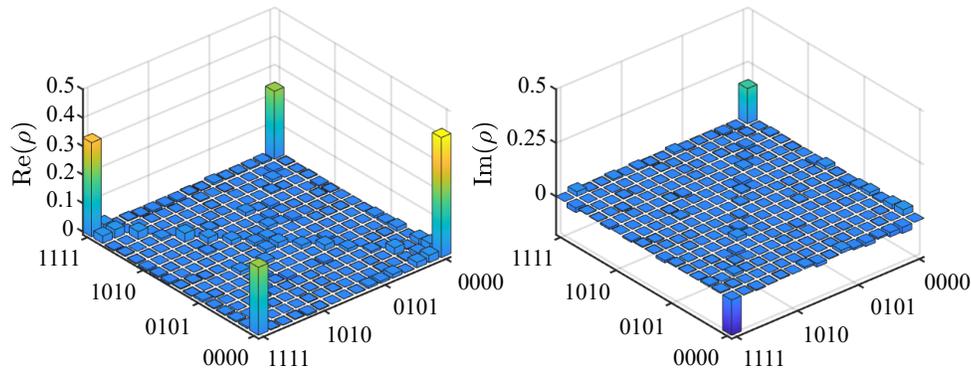

FIG. S3. Density matrix elements of the prepared four-qubit state $\rho$ obtained via Qiskit using the IBM Quantum Falcon Processor (ibmq_belem v1.0.3). The fidelity with respect to the GHZ state is estimated to be $78\%$.



\end{document}